  \documentclass[12pt,preprint]{aastex}%   
 \usepackage{natbib}   
 \usepackage{amsmath}   
 \usepackage{graphicx}%   
 \usepackage{amsfonts}%    
\usepackage{amssymb}   
\shorttitle{Viscosity- parameter and disk scale height}\shortauthors{Meirelles \& Leister} 
\begin{document}    
\title{The dependence of the viscosity- parameter 
on the disk scale height profile } 
\author{Cesar Meirelles Filho \& Nelson Leister}   
\affil{Instituto de Astronomia, Geof\'{\i}sica e de Ci\^{e}ncias Atmosf\'{e}ricas \\   
Universidade de S\~ao Paulo \\R. do Mat\~ao, 1226, 05508-090 S\~ao Paulo, SP, Brasil}
\email{cmeirelf@astro.iag.usp.br}   
\begin{abstract}   
	It is shown that the height scale for accretion disks is a constant whenever hydrostatic equilibrium and  sub-sonic turbulence regime hold in the disk. 
In order to have a variable height scale,  processes   that do contribute with  an extra term to the continuity equation are needed. This makes the viscosity parameter 
 much greater in the outer region and much smaller in the inner region. Under these circumstances, turbulence is a presumable source of viscosity in the disk.  
\end{abstract} 
\section{Introduction}
    In the last 35 years, accretion disks have become one of the most intense research areas in theoretical astrophysics. The reason for that  cannot be explained
 just by invoking their ubiquity in a variety of different astrophysical environment. At that point, one should recognize the role played by the  \citet{ss73} $\alpha$ standard model, 
which made appeal to a very intuitive, simple and neat physics. This model has, as its main assumptions, geometrical thinness, large optical thickness, 
 hydrostatic equilibrium in $z$, perpendicular to the plane of the disk, and the viscosity parametrization in terms of an unknown $\alpha$ parameter. It was meant 
 to apply under very specific conditions, such as those occurring in the outer parts of accretion disks in Bynary Systems, around very Young Stellar Objects, and in flows associated
 with the central engine for Active Galactic Nuclei \citep{palin95}. Soon, people realized that the approach used to treat viscosity was general and could be used 
under quite different conditions. Somehow this brought some ease to the long standing question of viscosity in accretion disks, encouraging people to tackle related problems, 
 even without knowing what the source of viscosity really is.  So successful and profuse the applications were that, today, one speaks of, at least, four models
 of accretion disks: the $\alpha$-standard model, successfully applied to Cataclysmic Variables, Transient X-Ray Sources, accretion disks around Active Galactic Nuclei, 
accretion disks in Young Stellar Objects (YSO) \citep{sma99, las01, men00, kin07, be85,lin89}; the ADAF model, where the acronym stands for Advection Dominated Accretion Flows,
 used to explain X and $\gamma$-ray emission from underluminous X-Ray Binaries and Active Galactic Nuclei \citep{per00, nayi95, bele03, naral02}; the 
CDAF model, Convection Dominated Accretion Flows, applied to underfed and  radiatively inneficient hard X-Ray Binaries and Active Galactic Nuclei
 \citep{abra02,naral00,igu02,igu03,naral02,quag00, ba01}; the  \citet{sha76}  two-temperature accretion disk model applied to Cygnus X-1.
Despite a remarkable difference among these models and systems to which they apply, the $\alpha$-parametrization works quite well. One of the reasons for that
 would be the weak dependence of the disk properties on $\alpha$ \citep{kin07}. Besides, it seems that the spread on the values of $\alpha$ to fit all these systems 
is not large, i.e., fairly similar values of $\alpha$ result in reasonable agreement with observations  \citep{kin07}. There are, however, some points that require 
more detailed treatment, and these are related to the disk scale height. By this we mean not only its value but, above all, how it behaves along the disk . 
 These questions are of fundamental importance when one is concerned with characteristic time scale lengths, turbulence regimes, and criteria to choose among
 different energy transport models. To make this point more clear, we shall focus on the $\alpha$-standard model, but our criticism applies to them all. 
The assumed constancy of the viscosity parameter is very decisive on the $\alpha$-standard model, leading to a disk scale height that goes like $r^{\frac{9}{8}}$, 
and yields the same behavior along $r$ for both the effective temperature and the temperature at the mid-plane ($z=0$) of the disk, $T \approx  r^{-{\frac{3}{4}}}$.
As a matter of fact, it leads to a constant optical depth along $r$. This result, apparently consistent, is made possible only by an unsound interpretation of a 
 formal solution to the continuity equation. Using the same assumptions of the $\alpha$-standard model, but the constancy of the viscosity parameter, and withdrawing the flaw
 by taking the correct solution to the continuity equation, the results change quite significantly. Now, $\alpha$ will go as $r^{11.25}$, with a huge variation along the disk.
 If constancy of $\alpha$ is indeed required, one will have to look for some process that add a term in the continuity equation in such a way as to give the correct 
 disk scale height. It is our claim in this letter that the questions we have just mentioned cast some doubts about the way the disk scale height is obtained  
and its implications in the determination of $\alpha$. The results we have obtained in this letter do modify previous results in the area, and do modify our 
 current knowledge about fundamental issues in accretion disk theory. Besides, they are very important as far as full consistency is required and have not been 
considered so far.

\section{The solution to the continuity equation and the radial height scale profile}
	In this section we are going to show how the radial scale height is related to the mass continuity equation. We also show that, in order to have a disk scale 
 height varying with radial distance, one needs to look for processes that add an extra term to the continuity equation.
    Since the disk is assumed to have azymuthal symmetry, being under hydrostatic equilibrium in z-direction, the continuity equation reads,   
 \begin{equation}    
{\frac{1}{r}}\, {\frac{\partial}{{\partial}\, r}}\, {\left(r\, \rho\, V_{r}\, \right)}=0\, ,\label{1}%    
\end{equation}
 with $r$, $\rho$, $V_{r}$ being, respectively, radial distance, mass density and radial velocity. It should be said that the above equation neglects mass transport 
due to turbulence, which is equivalent to assuming sub sonic turbulence regime.       Now, we set $f{\left(r\, , z\, \right)}= r\, \rho\, V_{r}\, $ and expand $f$ in powers of $z$ 
to obtain  
\begin{equation} 
f{\left(r\, , z\, \right)}=\sum_{n=0}^{\infty}\, \frac{{z}^{2\, n}}{{\left(2\, n\right)}!}\, {\left({\frac{{\partial}^{2\, n}}{{\partial}\, z^{2\, n}}}\, f{\left(r\, , z\, \right)}\right)}_{z=0}\,,\label{2}% 
\end{equation}
 where account was taken of reflexion symmetry over $z$.       Inserting $f{\left(r\, , z\, \right)}$ given by eq.(2) into eq.(1) gives  
\begin{equation} 
{\frac{1}{r}}\, \sum_{n=0}^{\infty}\, \frac{{z}^{2\, n}}{{\left(2\, n\right)}!}\, {\frac{\partial}{{\partial}\, r}}{\left({\frac{{\partial}^{2\, n}}{{\partial}\, z^{2\, n}}}\, f{\left(r\, , z\, \right)}\right)}=0\,.\label{3}%   
 \end{equation}
   Setting $z=0$, we must have  
 \begin{equation}  
 {\frac{\partial}{{\partial}\, r}}\, f{\left(r\, 0\right)}=0\,,\label{4}% 
 \end{equation}  
 or   
 \begin{equation} 
 f{\left(r\, , 0\, \right)} \equiv C_{0}\, ,\label{5}% 
 \end{equation}
 where $C_{0}$ is constant.    For $z \neq 0$, we recall that  
 \begin{equation}
{\frac{\partial}{{\partial}\, r}}\, {\left( {\frac{{\partial} ^{2\, n}}{{\partial} \,z^{2\, n}}}\, f{\left(r\, , z\, \right)}\right)}_{z=0}=
{\left( {\frac{{\partial} ^{2\, n}}{{\partial} \,z^{2\, n}}}\, {\frac{\partial}{{\partial}\, r}}\,f{\left(r\, , z\, \right)}\right)}_{z=0}\, ,\label{6}%    
\end{equation}
which, by eq.(1), should give
 \begin{equation}
 {\frac{{\partial} ^{2\, n}}{{\partial} \,z^{2\, n}}}\,{\frac{\partial}{\partial \, r}}\, f{\left(r\, , z\, \right)}=0\,. \label{7}% 
 \end{equation}
Then   
 \begin{equation}  
{\left({{\frac{{\partial} ^{2\, n}}{{\partial} \,{z^{2\, n}}}}\, f{\left(r\, , z\, \right)}}\right)}_{z=0}\equiv constant\equiv C_{2\, n}\, ,\label{8}%  
 \end{equation}
 and the most general solution to eq.(1) should be written 
 \begin{equation}
 f{\left(r\, , x\, \right)}=C_{0}\, \sum_{n=0}^{\infty} \frac{C_{2\, n}}{C_{0}\, {\left(2\, n \right)}\, !}\,x^{2n}\,, \label{9}% 
 \end{equation} 
where $x={\frac{z}{\ell}}$, and $C_{0}={\left( r\, \rho\, V_{r}\, \right)}_{z=0}$.       Finally, we may write for the accretion rate,  
\begin{equation} 
{\dot M}=4\, \pi\, \ell\, C_{0}\, \sum_{n=0}^{\infty} {\frac{C_{2\, n}}{C_{0}\, {\left(2\, n+1\right)}\, !}}\, ,\label{10}%   
\end{equation}
and since   
\begin{equation} 
{\frac{\partial}{{\partial}\, r}}\, {\dot M}=0\, ,\label{11}%   
\end{equation}  
this implies   
\begin{equation}  
{\frac{\partial}{{\partial}\, r}}\, {\ell}=0\, ,\label{12}%  
\end{equation} 
because the $C_{2\, n}$ are all constants.  
     We, then, must conclude that, under hydrostatic equilibrium together with sub sonic turbulence, the height scale of the disk is not allowed to vary along 
the radial distance $r$.  

\section{The dependence of the viscosity parameter on the disk height scale}   
    	In this section, we are going to highlight some consequences of the conclusions we have drawn in the last section. In order to proceed, let us recall some results
very familiar from the accretion disks theory. Let us start from the hydrostatic equilibrium equation,  
\begin{equation}  
\frac{\partial}{{\partial}\, z}\, P=-\rho\, {\Omega}^{2}\, z\, ,\label{13}%  
\end{equation}
which serves to define the disk height scale $\ell$, i.e., 
\begin{equation} 
P=\rho\, {\Omega}^{2}\, {\ell}^{2}\, ,\label{14}%  
\end{equation}  
with  $P$ being the pressure, and $\Omega$ the Keplerian angular velocity.
    	The angular momentum conservation equation may be written  
\begin{equation} 
{\dot M}\, \Omega\, r=4\, \pi\, {\frac{\partial}{{\partial}\, r}}{\left( \alpha\, P\, {\ell}^{2}\, r\, \right)}\, ,\label{15}%   
\end{equation} 
from which we obtain
\begin{equation} 
\rho={\frac{\dot M}{2\, \pi\, \alpha\, \Omega\, {\ell}^{3}}}\, S\, ,\label{16}%
\end{equation}
$\alpha$ being the viscosity parameter, and  
\begin{equation} 
S=1-{\left( {\frac{r_{1}}{r}}\right)}^{\frac{1}{2}}\, \label{17}% 
\end{equation}
takes into account null boundary condition for the torque at $r=r_{1}$, the inner radius of the disk.       Assuming the disk to be cooled by radiative transport in $z$
 direction, we may write, in the diffusion approximation, 
\begin{equation}
F_{z}=-c\, {\frac{\partial}{{\partial}\, \tau}}\, P_{r}\, ,\label{18}%   
\end{equation}
with $c$, $F_{z}$, $\tau$, $P_{r}$ being, respectively, velocity of light, radiative flux in $z$ direction, optical depth, and radiation pressure. Replacing differentials by 
finite differences, and recalling the definition of effective temperature, i.e., 
\begin{equation} 
\sigma\, {T_{eff}}^{4}={\frac{3}{4\, \pi}}\, {\dot M}\, {\Omega}^{2}\,,\label{19}%  
\end{equation} 
eq.(18) may be written as 
\begin{equation}
\frac{3}{4\, \pi\, \sigma}\, {\dot M}\, {\Omega}^{2}\, \tau=T^{4}\, ,\label{20}% 
\end{equation}
with  $\sigma$ being the Stefan-Boltzmann constant, and $T$ the temperature at the mid plane of the disk.       In the outer parts of the disk, the opacity is mainly given 
by the free-free opacity.So, using a Rosseland mean opacity, averaged over $z$, eq.(20) will be rewritten as
\begin{equation} 
T^{4}=2.62\times 10^{26}\, {\dot M}\, {\Omega}^{2}\, {\rho}^{2}\, T^{-3.5}\, {\ell}\, .\label{21}%
\end{equation}
  	Finally, for a gas pressure dominated disk, we obtain from eq.(21), 
\begin{equation}
{\alpha}^{2}=1.33\times 10^{-51}\, \frac{{\dot M_{17}}^{3}}{M_{34}}\, \frac{x^{22.5}}{y^{20}}\, ,\label{22}% 
\end{equation} 
where ${\dot M_{17}}$, ${M_{34}}$, $x$, $y$ are, respectively, accretion rate in units of $10^{17}\, g\, s^{-1}$, mass of the central object in units of
 $10^{34}\, g$, the radial distance in units of the inner radius, the disk scale height in units of the inner radius. The inner radius $r_{1}$ is assumed to be $3\, R_{g}$,
 $R_{g}$ being the gravitational radius.  
	According to \citet{kin07}, if constraints on $\alpha$ are required, one should make resource to observations of systems subject to temporal behavior, such as 
 dwarf nova outbursts \citep{war03, can01}, outbursts of X-ray transients \citep{las01}, protostellar accretion disks \citep{har98}, FU Orionis outbursts \citep{loclar04}, 
 or optical variability in  Active Galactic Nuclei \citep{star04}. If we take, for instance, the viscous time-scale,
\begin{equation}
t=-{\int_{r}^{r_{1}}}\,{\frac{r}{\nu}}\,dr \, ,\label{24}% 
\end{equation}
where $\nu$ is the kinematic viscosity, we obtain for the standard model after little algebra, using eq.(22),
\begin{equation}
\frac{\ell}{r}=\frac{y}{x}=1.63\times 10^{-3}\, {\frac{{\dot M_{17}}^{0.225}}{{M_{34}}^{0.325}}}\, x_{d}^{-0.319}\, t^{0.25}\, x^{0.125}\, ,\label{24}%
\end{equation}
$x_{d}$ being the disk size in our units.
	Now, inserting this expression into eq.(22), it yields
\begin{equation}
{\alpha}_{s}=207.53\,{ M_{34}}^{1.375}\, {\dot M_{17}}^{-0.375}\, x_{d}^{1.575}\, t^{-1.25}\, ,\label{25}%
\end{equation}
or, to make a comparison with \citet{sma99},
\begin{equation}
t=71.39 \,{M_{34}}^{1.1}\, {{\dot M}_{17}}^{-0.3}\, {{\alpha}_{s}}^{-0.8}{x_{d}}^{1.25}\, .\label{26}%
\end{equation} 
	It should be remarked that, contrary to \citet{sma99} paper, we use the disk scale height dependence on $T$, the temperature at $z=0$, not $T_{eff}$, 
which we believe to be the correct procedure. Essentially, this makes our scale height greater by a factor of ${\tau}^{0.125}$, $\tau$ being the optical depth. 
Our results differ from \citet{sma99} by minor discrepancies in the exponents, but by a numerical factor of about two orders of magnitude greater.
	Finally, using our formulation, we obtain
\begin{equation}
\frac{\ell}{r}=\frac{y}{x}=2.1\times 10^{-3}\, {\left(\frac{{\dot M}_{17}}{M_{34}}\right)}^{0.1875}\, t^{0.125}\, x^{-1}\, ,\label{27}%
\end{equation}
and for ${\alpha}_{0}$, the value of $\alpha$ directly related to the viscous time,
\begin{equation}
{\alpha}_{0}=0.1\, {x_{d}}^{-1.575}\, {\alpha}_{s}\, , \label{28}%
\end{equation}
with ${\alpha}_{s}$ given by eq.(25). Expression (28) is the value of $\alpha$ at $x=1$. The value of $\alpha$ anywhere in the disk is
\begin{equation}
\alpha=x^{11.25}\, {\alpha}_{0}\, .\label{29}%
\end{equation}

	From eq.(25) and eq.(28) we see that the viscosity parameter related to the viscous time will be much smaller when we employ the correct solution to the 
continuity equation. At the outer radius of the disk it will be $0.1\, {x_{d}}^{9.675}$ greater than the value obtained with the standard model.

\section{Analysis and conclusions}   

 A rapid inspection on eq.(22) reveals a very strong dependence of $\alpha$ on the radial distance, due to the assumptions of sub sonic
 turbulence and hydrostatic equilibrium. The $\alpha$ parametrization of the viscosity yields hardly credible results. In a disk of $x_{d}=100$, $\alpha$ varies by 
a factor of $3\times 10^{22}$ as compared to its value at $x=1$. Assuming equality between the length scale of the turbulence and the height scale of the disk, since the disk 
is assumed to be thin, we have for $y=0.01$, sub sonic turbulence only for $x\leq 3.04$; and for $y=0.1$, only for $x\leq 23.56$. The results are highly dependent on the extent and
 thinness of the disk. The thicker the disk, the more sub sonic it will be.    It should be stressed that the conclusions we have drawn are based on the analysis of a solution obtained under
 conditions that, under the usual procedure, would result in the $\alpha$ standard model, which gives a disk scale height varying with $r^{\frac{9}{8}}$, a necessary condition 
needed to have constant $\alpha$. However, the constancy of $\ell$ along $r$ is not an assumption, but it stems as a consequence from a rigorous solution to the mass continuity 
equation. In other words, the assumption of constant $\alpha$ is not compatible with the solution to the mass continuity equation. To make things compatible, we should have
\begin{equation}
{\frac{1}{r}}\, {\frac{\partial}{{\partial}\, r}}\, {\left(r\, \rho\, V_{r} \right)}+{\it L}{\left(\rho\, , {\vec V}\right)}=0\, ,\label{26}%   
\end{equation}
 where $\it L=L{\left({\frac{\partial}{{\partial}\, r}}\,,{\frac{\partial}{{\partial}\, z}}\,\right)}$ is an operator applied to $\rho$ and ${\vec V}$. If we insist in hydrostatic equilibrium, $\it L$ describes turbulent mass transport.       It is beyond our aim, in this
 letter, to go any further on this matter but, if a disk height scale dependence on $r$ is essential to have physically meaningful results, a urgent search for physical processes
 that do contribute with an extra term to the mass transport equation is indeed required. In that respect, it is very unlikely to discard turbulent mass transport in the disk
 as one of the reasons to have the disk scale height varying along the radial distance, and, if so, turbulence is a presumable source of viscosity in the disk.
	The points we have raised in this letter deserve attention by themselves but in no way exhaust the subject. The value of the disk height scale and the way it behaves along $r$
 are decisive to know, among other things, what is the relevant energy transport mechanism in a given region and how it varies along the disk. It is our intention to
 to consider these issues, in a more detailed way, in a future article.

\acknowledgments                         

The authors  acknowledge support from FAPESP .

%%%%%%%%%%%%%%%%%%%%%%%% end refs. %%%%%%%%%%%%%%%%%%%%%%%%%%%%%%%%%%%%%%

%%%%%%%%%%%%%%%%%%%%%%%%% figuras   %%%%%%%%%%%%%%%%%%%%%%%%%%%%%%%%%%%%%%

\clearpage                                    

\end{document}